\documentclass[a4paper]{panl}
\usepackage{cite}
\usepackage{wrapfig}
\usepackage{graphicx}
\usepackage{amssymb}
\usepackage{amsfonts}
\usepackage{amsmath}
\usepackage{longtable}
\usepackage{rotating}
\usepackage{lscape}
\usepackage{epsfig}
\usepackage{multirow}

\newcommand{\pder}[2]{\frac{\partial #1}{\partial #2}}
\newcommand{\dpder}[2]{\frac{\partial^2 #1}{\partial {#2}^2}}
\newcommand{\grad}[1]{{\, \text {grad} #1}}
\renewcommand{\div}[1]{\, \text {div} #1}
\newcommand{\rot}[1]{\, \text {rot} #1}

\usepackage{lineno}

\originalTeX
\begin{document}

\title{Resonant generation of high-order harmonics in
nonlinear electrodynamics}
\maketitle
\authors{I.\,Kopchinskii$^{a,b,}$\footnote{E-mail: kopchinskii@ms2.inr.ac.ru},
P.\,Satunin$^{b,a}$\footnote{E-mail: satunin@ms2.inr.ac.ru}}
\from{$^{a}$\,Lomonosov Moscow State University, Faculty of Physics}
\from{$^{b}$\,Institute for Nuclear Research of RAS}

\begin{abstract}

We study the process of resonant generation of high-order harmonics in a closed cavity in the model of vacuum nonlinear electrodynamics. Concretely, we study the possibility of resonant generation of the third harmonic induced by a single electromagnetic mode in a radiofrequency cavity, as well as resonant generation of  a combined frequency mode induced by two pump modes ($\omega_1$ and $\omega_2$).
We explicitly show that the third harmonic as well as the $2\omega_1+\omega_2$ combined frequency mode are not resonantly amplified, while the $2\omega_1-\omega_2$ signal mode is amplified for certain cavity geometry. We discuss the process from the point of view of quantum theory.
\end{abstract}
\vspace*{6pt}

\label{sec:intro}
\section*{Introduction}

Vacuum nonlinearity is one of phenomena theoretically predicted at the dawn of quantum electrodynamics\cite{Euler:1935zz,Heisenberg:1935qt} but have not experimentally detected yet due to its extreme smallness.  A part of the effects of vacuum nonlinearity related to electrodynamics mimic to similar effects in nonlinear optical crystals, which include vacuum birefringence of a photon in external field, and the generation of high-order harmonics. The latter can be tested both in optical and radio wave range. The generation of high-order harmonics for radio modes in cavity may be potentially detected only in case of resonant increase \cite{Brodin:2001zz}, see partial solutions in \cite{Eriksson:2004cz, Bogorad:2019pbu}. An interesting issue is to study this nonlinear process for arbitrary set of cavity mode, which can be done analytically for rectangular shape of cavity. Thus, studying  nonlinear effects for single mode in one-dimensional cavity it turns out that there is no resonant generation of the third harmonics \cite{Shibata:2020don}. We generalise the method to the case of two pump modes and 3D rectangular cavity.  More details  are given in \cite{OurArticle}; in addition to that article we discuss the quantum aspects of the obtained result in the conclusion.

\newpage
\label{sec:preparation}
\section*{General theory for resonance}

We start from Euler-Heisenberg effective Lagrangian \cite{Euler:1935zz,Heisenberg:1935qt},
\begin{equation}
\label{eq:lagrangian}
\mathcal{L} = -\frac{1}{4} \mathcal{F} + \kappa \, (\mathcal{F}^2 + \beta \, \mathcal{G}^2), \qquad \kappa = \frac{\alpha_e^2 }{90 \, m_e^4}, \quad \beta = \frac{7}{4},
\end{equation}
 The electromagnetic field invariants have the standard form, 
\begin{equation}
\label{eq:invariants}
\mathcal{F} \equiv F_{\mu\nu} F^{\mu\nu} = -2\left(\textbf{E}^2 -  \textbf{B}^2\right), \qquad\qquad \mathcal{G} \equiv F_{\mu\nu} {\widetilde F}^{\mu\nu} = -4\left(\textbf{E} \cdot \textbf{B} \right).
\end{equation}
Varying the Lagrangian, one obtains modified Maxwell equations \cite{Euler:1935zz}:
\begin{equation}
\label{eq:mod-maxwell}
\begin{aligned}
\rot{\textbf{B}} &=  \pder{\textbf{E}}{t} + \left[\pder{\textbf{P}}{t} - \rot{\textbf{M}}\right], \\
\rot{\textbf{E}} &= -\pder{\textbf{B}}{t}, \\
\end{aligned}
\hspace{2cm}
\begin{aligned}
\div{\textbf{B}} &= 0, \\
\div{\textbf{E}} &= -\div{\textbf{P}},
\end{aligned}
\end{equation}
where  $\textbf{P}$ and $\textbf{M}$ denote vacuum polarization and magnetization respectively,
\begin{equation}
\label{eq:corrections}
\begin{aligned}
\textbf{P}(\textbf{x}, t) &\equiv 16 \, \kappa \left[\left(\textbf{E}^2 -  \textbf{B}^2\right) \textbf{E} + 2 \beta \left(\textbf{E} \cdot \textbf{B} \right) \textbf{B}\right], \\
\textbf{M}(\textbf{x}, t) &\equiv 16 \, \kappa \left[\left(\textbf{E}^2 -  \textbf{B}^2\right) \textbf{B} - 2 \beta \left(\textbf{E} \cdot \textbf{B} \right) \textbf{E}\right].
\end{aligned}
\end{equation}
Wave equations both for amplitudes for electric and magnetic fields are modified as well \cite{Euler:1935zz},
\begin{equation}
\label{eq:mod-waves}
\begin{aligned}
\Box \textbf{E} &= \pder{}{t} \rot{\textbf{M}} +  \grad{\div{\textbf{P}}} - \dpder{\textbf{P}}{t} , \\
\Box \textbf{B} &=  \pder{}{t} \rot{\textbf{P}} - \grad{\div{\textbf{M}}} + \Delta \textbf{M} .
\end{aligned}
\end{equation}
These equations are nonlinear; the solving  them seems to be a hard issue. However, for the our goal we may apply the perturbation theory. 

We call the mode, initially given in the cavity, as a ``pump mode'' (electric field $\textbf{E}^p$),  and look for the evolution of ``signal mode'' (electric field $\textbf{E}^{sig}$) which was not initially present in the cavity.  Assuming the hierarchy $|\textbf{E}^{sig}| \sim \kappa \left( |\textbf{E}^{p}|\right)^3 \ll |\textbf{E}^{p}|$, 
one obtains in the zeroth order homogeneous  wave equations for the pump modes $\Box \textbf{E}^{p} = 0, \ \Box \textbf{B}^{p} = 0$, and in the first order inhomogeneous  linear wave equations for signal mode amplitudes:
\begin{equation}
\label{eq:linear}
\begin{aligned}
\Box \textbf{E}^{sig} &= \pder{}{t} \rot{\textbf{M}(\textbf{E}^{p}, \textbf{B}^{p})} + \grad{\div{\textbf{P}(\textbf{E}^{p}, \textbf{B}^{p})}} - \dpder{\textbf{P} (\textbf{E}^{p}, \textbf{B}^{p})}{t} , \\
\Box \textbf{B}^{sig} &= \pder{}{t} \rot{\textbf{P}(\textbf{E}^{p}, \textbf{B}^{p})} - \grad{\div{\textbf{M}(\textbf{E}^{p},\textbf{B}^{p})}} + \Delta \textbf{M}(\textbf{E}^{p}, \textbf{B}^{p}) .
\end{aligned}
\end{equation}
These inhomogeneus wave equations may have resonantly growing solutions if the r.h.s. contains terms which coincide with the solution of corresponding homogeneous equations. This resonance means the linear growing with time (see 1D component for example),
$$
\Box E_x = \cos (\omega_r t)\sin (\omega_r x)\,+\,...  \ \ \to \ \ E_x^{(growing)} = \frac{t}{\omega_r}\sin (\omega_r t) \sin (\omega_r x).
$$
In a real world this linear growth stops when the dissipation effect become significant. Introducing dissipation coefficient $\Gamma$ ``by hands'', on obtains the saturation of the linear growth:
$$
(\Box + \Gamma \partial_t) E_x = \cos (\omega_r t)\sin (\omega_r x)  \ \ \to \ \ E_x^{(steady)} = \frac{1}{\Gamma \omega_r}\sin (\omega_r t) \sin (\omega_r x),
$$
the amplitude of resonant mode inverse proportianal to $\Gamma$.

We apply this approach to single and two pump modes in 1D and 3D cavities.

\section*{One-dimensional cavity}
\begin{figure}[ht!]
\centering
\includegraphics[width=0.9\linewidth]{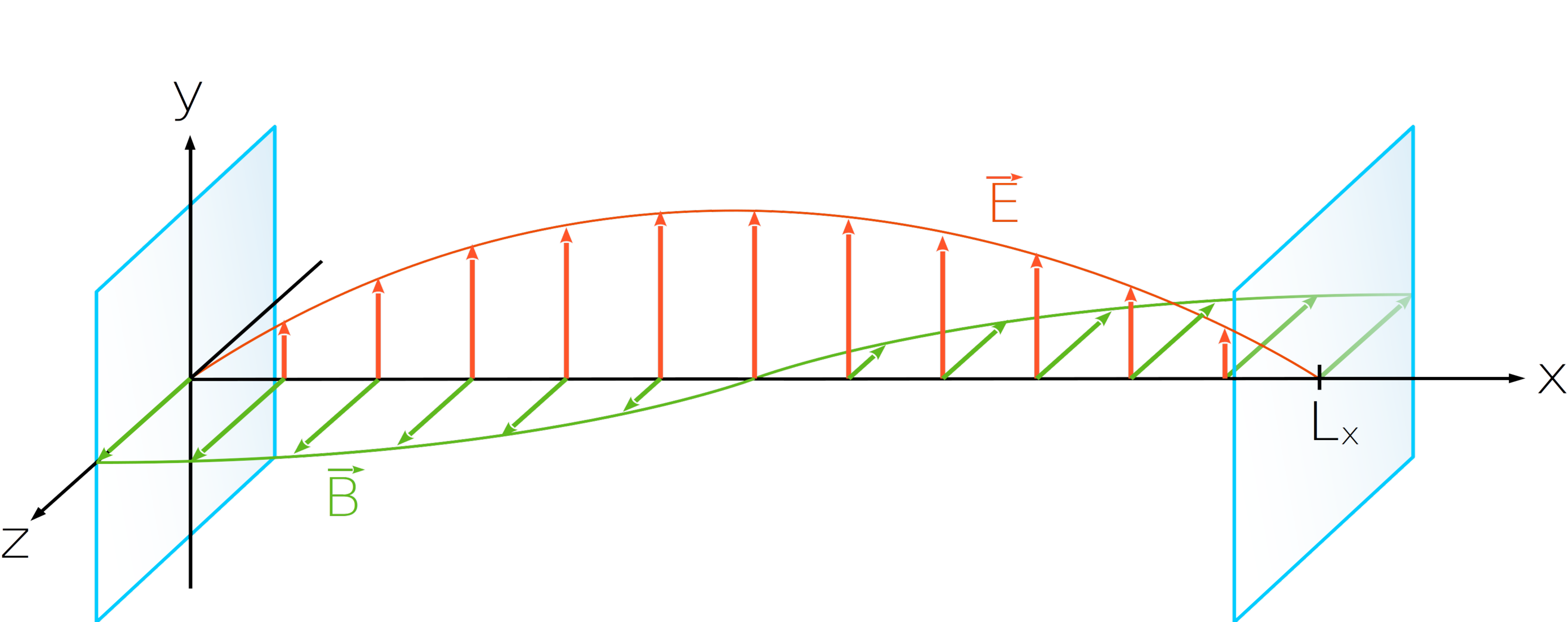}
\caption{Field configuration of an electromagnetic mode in one-dimensional cavity ($n=1$).}
\label{Fig:1D}
\end{figure}

We call by ``one-dimensional cavilty''  a rectangular cavity with  one spatial dimension smaller than others, say $L_x \ll L_y, L_z$. A pump mode configuration reads (only nonzero components),
\begin{equation}
\label{1ddpump}
E^{p}_y(x, t) = F_0 \cdot \sin(k_n x) \sin(\omega_n t), \qquad 
H^{p}_z(x, t) = F_0 \cdot \cos(k_n x) \cos(\omega_n t),
\end{equation}
where the eigenfrequencies and wave vectors take discrete number of values: $\omega_n = k_n = n\cdot \frac{\pi}{L_x}, \ n \in \mathbb{N}$. The corresponding electromagnetic field configuration is shown at Fig.~\ref{Fig:1D}.

To study the evolution of a signal mode initiated by one pump mode in cavity, we write the linearized wave equations (\ref{eq:linear}) with r.h.s. calculated at the pump mode configuration (\ref{1ddpump})\footnote{Analytical calculations were made in ``wxMaxima 21.02.0'' computer algebra system \cite{Maxima}} (cf. \cite{Shibata:2020don}), 
\begin{equation}
\footnotesize
\label{single1d}
\begin{aligned}
\left(\Box + \Gamma\partial_t\right) E_y^{sig} &= 8\kappa F_0^3\omega_n^2 \Bigl[2\sin({\omega_n} x)\sin({\omega_n} t) - 3\sin({\omega_n} x)\sin({3\omega_n} t) + \sin({3\omega_n} x)\sin({\omega_n} t)\Bigr],
\\
\left(\Box + \Gamma\partial_t\right) H_z^{sig} &= 8\kappa F_0^3\omega_n^2 \Bigl[2\cos({\omega_n} x)\cos({\omega_n} t) - \cos({\omega_n} x)\cos({3\omega_n} t) + 3\cos({3\omega_n} x)\cos({\omega_n} t)\Bigr].
\end{aligned}
\end{equation}
The first terms in both lines of the r.h.s. of eqs.~(\ref{single1d}) represent solution of homogeneous equation so there is a resonant mode with the pump mode frequency $\omega_n$; the other two terms are non-resonant. Surprisingly, there are no resonant terms with triple frequency in the r.h.s of eqs.~(\ref{single1d}) so the third harmonics of the pump mode does not appear.

For the simplicity of presentation of the following results we present the r.h.s. as a table, see Table~\ref{tbl:single-1D}.
\begin{table}[ht!]
\centering
\def\arraystretch{1.3}
\begin{tabular}{|l|c|c|}
    \hline
   wavenumbers & $n$ & $3n$ \\
    \hline
    eigenfrequencies& $\omega_{n}, ~ \omega_{3n}$ & $\omega_{n}$ \\
    \hline
\end{tabular}
\caption{Examination of the resonance criterion for a single pump mode in 1D-cavity.}
\label{tbl:single-1D}
\end{table}
We see that the only eigenfrequency connected with wave vector is $\omega_n$.

The next step is to consider two pump modes in one-dimensional cavity. Generally, we should take into account arbitrary angle $\alpha$ between polarization plane of two modes with wavenumbers $n$ and $p$. Following the same algorithm as in the the previous case, we calculate the r.h.s. on the pump mode combination which is shortly presented as Table~\ref{tbl:two-1D} (symmetric part with $n \leftrightarrow p$ omitted).
\begin{table}[ht!]
\centering
\def\arraystretch{1.3}
\begin{tabular}{|l|c|c|c|c|}
    \hline
    wavenumbers & ${n}$ & $3n$ & ${2n-p}$ & ${2n+p}$  \\
    \hline
    eigenfrequencies & ${\omega_{n}}, ~ \omega_{2p\pm n}, ~ \omega_{3n}$ & $\omega_{n}$ & $\omega_{p}, ~ {\omega_{2n+p}}$ & $\omega_{p}, ~ {\omega_{2n-p}}$\\
    \hline
\end{tabular} 
\caption{Examination of the resonance criterion for two pump modes in 1D-cavity.}
\label{tbl:two-1D}
\end{table}
We see that the only resonant modes have wavenumber $n$; the mixed signal modes with wavenumbers $2n \pm p$ do not resonate.

\section*{3D rectangular cavity.}

The next step is to consider one or two modes in 3D rectangular cavity of dimensions ($L_x,L_y,L_z$). The cavity modes are classified as $TE_{npq}$ and $TM_{npq}$ modes, see \cite{Hill:2014}, where $n,p,q$ numerate wavenumbers for each spatial dimension. The wavevector for each mode is determined as $\textbf{k}_{npq} = \left(\frac{\pi n}{L_x},\, \frac{\pi p}{L_y}, \, \frac{\pi q}{L_z}\right)$, the same for $TE$ and $TM$ modes; the corresponding frequency is
\begin{equation}
\label{Disp}
\omega_{npq} = | k_{npq}|= \pi \sqrt{\frac{n^2}{L_x^2} + \frac{p^2}{L_y^2} + \frac{q^2}{L_z^2}}.
\end{equation}
First, consider single mode in the cavity, and apply  aforementioned algorithm in WxMaxima for resonant signal mode searching. The result is shown in Table 3 
.
\begin{table}[ht!]
\centering
\def\arraystretch{1.3}
\begin{tabular}{|l|c|c|c|c|c|}
    \hline
    wavenumbers & $n,p,q$ & $3n,p,q$ & $3n,3p,q$ & $\left(n \leftrightarrow p \leftrightarrow q\right)$ & $3n,3p,3q$ \\
    \hline
    eigenfrequencies  & \multicolumn{4}{|c|}{$\omega_{npq}, ~ 3\omega_{npq}$} & $\omega_{npq}$ \\
    \hline
\end{tabular}
\label{tbl:single-3D}
\caption{Examination of the resonance criterion for a single pump mode in 3D-cavity.}
\end{table}
The only resonant mode has  the same frequency as the pump mode; the third harmonics does not resonantly amplified.

Continue with the two pump mode configuration. This is the most difficult case due to complicated analytical calculations even using computer algebra system. First, due to trigonometric relations the r.h.s. of  eqs.~(\ref{single1d}) contain terms only of the form:
$A\,h(\omega_{sig} t)\,h(k_{sig,x} x)\,h(k_{sig,y} y)\,h(k_{sig,z} z),$ where the function $h()$ means sine or cosine, and $\omega_{sig}$ and $k_{sig}$ take the values of the following sets,
\begin{equation}
\label{eq:combs}
\begin{matrix}
\hspace{2mm}\omega_{sig} ~\in~ \{ & \omega_1, \quad & \omega_2, \quad & 2\omega_1 \pm \omega_2, \quad & 2\omega_2 \pm \omega_1, \quad & 3\omega_1, \quad & 3\omega_2 \quad \hspace{1mm}\}, \\
k_{sig,x} ~\in~ \{ & k_{1x}, \quad & k_{2x}, \quad & 2k_{1x} \pm k_{2x}, \quad & 2k_{2x} \pm k_{1x}, \quad & 3k_{1x}, \quad & 3k_{2x} \quad\}, \\
k_{sig,y} ~\in~ \{ & k_{1y}, \quad & k_{2y}, \quad & 2k_{1y} \pm k_{2y}, \quad & 2k_{2y} \pm k_{1y}, \quad & 3k_{1y}, \quad & 3k_{2y} \quad\}, \\
k_{sig,z} ~\in~ \{ & k_{1z}, \quad & k_{2z}, \quad & 2k_{1z} \pm k_{2z}, \quad & 2k_{2z} \pm k_{1z}, \quad & 3k_{1z}, \quad & 3k_{2z} \quad\}. \\
\end{matrix}
\end{equation}
The r.h.s. of the linearized wave equations (\ref{eq:linear}) generally contain terms from different columns of (\ref{eq:combs}), thus the mixed terms like $\sin(3\omega_1 t)\cos[(2k_{1x}-k_{2x})x]\cos(k_{1y}y)\sin(3k_{2z}z)$ might hypothetically appear.  However, the corresponding amplitude may vanish, as for the third harmonics. As previously, we calculate the r.h.s. in WxMaxima system, and obtain the result which can be presented in Table 4. The terms of the Table 4 can be grouped into two sectors: the triple wavenumbers are not associated with combined frequencies $2\omega_1 \pm \omega 2$ (means the corresponding amplitude vanishes) while the combined wavenumbers are not associated with triple frequencies. In addition, it turns out that  the amplitude vanishes for parallel wavevectors, $\textbf{k}_1 \parallel \textbf{k}_2$. 
\begin{table}[ht!]
\centering
\def\arraystretch{1.3}
\begin{center}
\begin{tabular}{|*{6}{c|}}
    \hline
    wavenumbers & $\begin{array}{c} n_1 \\ p_1 \\ q_1 \end{array}$ & $\begin{array}{c} 3n_1 \\ p_1 \\ q_1 \end{array}$ & $\begin{array}{c} n_1 \\ 3p_1 \\ q_1 \end{array}$ & $\begin{array}{c} \  \\ \cdots \\ \  \end{array}$ & $\begin{array}{c} 3n_1 \\ 3p_1 \\ 3q_1 \end{array}$ \\
    \hline
    eigenfrequencies& $\omega_1, ~ 3\omega_1, ~ 2\omega_2+\omega_1, ~ 2\omega_2-\omega_1$ & \multicolumn{3}{c|}{$\omega_1, ~ 3\omega_1$} & $\omega_1$ \\
    \hline
\end{tabular}
\begin{tabular}{|*{6}{c|}}
    \hline
    wavenumbers & $\begin{array}{c} 2n_2 \pm n_1 \\ p_1 \\ q_1 \end{array}$ & $\begin{array}{c} n_1 \\ 2p_2 \pm p_1 \\ q_1 \end{array}$ & $\begin{array}{c} n_1 \\ p_1 \\ 2q_2 \pm q_1 \end{array}$ & $\begin{array}{c} \cdots \end{array}$ & $\begin{array}{c} 2n_2 \pm n_1 \\ 2p_2 \pm p_1 \\ 2q_2 \pm q_1 \end{array}$  \\
    \hline
   eigenfrequencies & \multicolumn{5}{c|}{$\omega_1, ~ 2\omega_2+\omega_1, ~ 2\omega_2-\omega_1$} \\
    \hline
\end{tabular} 
\end{center}
\label{tbl:two-3D}
\caption{Examination of the resonance criterion for  two pump modes in 3D-cavity.}
\end{table}

Let us prove that the resonant generation does not appear for signal modes with frequency $2\omega_2 + \omega_1$. Write the dispersion relation for this signal mode and apply the triangle inequality,
\begin{equation*}
    2\omega_2 + \omega_1 = |2\textbf{k}_2| + |\textbf{k}_1| \geqslant |2\textbf{k}_2 + \textbf{k}_1|= \sqrt{\sum_{i=1}^3(2k_{2,i}+k_{1,i})^2}.
\end{equation*}
In case of non-parallel wavevectors the triangle inequality holds if at least for one $i$ $k_{sig,i} = 3\times \max \left(k_{1,i},~k_{2,i}\right)$ --- explicitly the case for which the amplitude vanishes.  
Thus, we show that the signal mode with frequency
 $2\omega_2 + \omega_1$ does not appear.

\section*{Resonant solution for $2\omega_1 - \omega_2$.}

In this section we show explicitly the resonant solution describing the generation of the combined mode with frequency $2\omega_1-\omega_2$. Consider modes $TE011$ and $TM110$ for concreteness. Table 4 shows two possible options for the signal mode: $2\omega_{011}-\omega_{110} = \omega_{130}$ and $2\omega_{011}-\omega_{110} = \omega_{132}$. Checking additionally the dispersion relation (\ref{Disp}) for each mode, we came that the second option cannot be realised for any cavity dimensions, but the first indeed can. The following condition on the ratio of cavity dimensions reads,
$$
  \left(\frac{L_z}{L_x}\right)^2 \left(\frac{L_z}{L_y}\right)^2 + \left(\frac{L_z}{L_x}\right)^2 +3\left(\frac{L_z}{L_y}\right)^2 = 1.
$$
Assuming additionally $L_x = L_y$, one obtains the resonant ratio of cavity dimensions $L_x:L_y:L_z = 1:1:r, \quad r = \sqrt{\sqrt{5} - 2} \approx 0.486$. Resonant conditions for other set of modes are expected to be computed similarly.

The corresponding signal amplitude, even being resonantly amplified, is rather small but detectable by single-photon detectors, see \cite{Bogorad:2019pbu,OurArticle} for details.

\section*{Summary and Discussion.}

We summarize the conclusions as follows.
First, we have shown explicitly that the third harmonic is not resonantly amplified both in 1D and 3D
rectangular cavities for arbitrary pump mode. Combined ``plus'' harmonic (frequency $2\omega_1 + \omega_2$) is not resonantly amplified in 1D and
3D rectangular cavities for arbitrary set of pump modes. From the other hand, the ``minus'' mode ($2\omega_1 - \omega_2$) is resonantly amplified in 3D cavity of certain resonant ratio of
dimensions. 

Let us have one more glance on the process $2\omega_{011}-\omega_{110} \to \omega_{130}$. Note that the wavevector components combinate independently  of each over with $\pm$ signs, the wavenumbers $(130)$ of the signal mode results as follows,  $(1=0+0+1)$, $(3= 1+1+1)$, $(0 = 1-1+0)$.

From the point of view of quantum theory, the generation of the signal mode is associated with the process  $3 \to 1$, the merging of three quanta of cavity modes into a single one. At first glance it seems to be a contradiction: the only possible result of such process is the energy of final quanta $2\omega_1 + \omega_2$ due to the energy conservation, while the classical approach lead to the energy $2\omega_1 - \omega_2$.

This apparent contradiction can be explained as follows. The classical waves (and cavity modes) are the coherent states. Each coherent state decomposes to the linear combination of the states with definite number of particles; nonzero contribution to the process are given from the amplitudes like
$$
|N_{\omega_1}, M_{\omega_2} \rangle \to | (N-2)_{\omega_1} (M+1)_{\omega_2} 1_{2\omega_1 - \omega_2} \rangle,
$$
for arbitrary values $M$ and $N$. The calculation of this amplitudes in a pure quantum approach will be presented in a following paper.

\paragraph{Acknowledgments} The Authors thank Maxim Fitkevich, Dmitry Kirpichnikov, Dmitry Levkov, \fbox{Valery Rubakov}, Alexey Rubtsov and Dmitry Salnikov for helpful discussions. The work is supported by RSF grant 21-72-10151.

\bibliographystyle{pepan}
\bibliography{pepan_biblio}

\begin{thebibliography}{1}
\def\selectlanguageifdefined#1{
\expandafter\ifx\csname date#1\endcsname\relax
\else\selectlanguage{#1}\fi}
\providecommand*{\href}[2]{{\small #2}}
\providecommand*{\url}[1]{{\small #1}}
\providecommand*{\BibUrl}[1]{\url{#1}}
\providecommand{\BibAnnote}[1]{}
\providecommand*{\BibEmph}[1]{\emph{#1}}
\ProvideTextCommandDefault{\cyrdash}{\hbox to.8em{--\hss--}}
\providecommand*{\BibDash}{\ifdim\lastskip>0pt\unskip\nobreak\hskip.2em\fi
\cyrdash\hskip.2em\ignorespaces}

\bibitem{Euler:1935zz}
\selectlanguageifdefined{english}
\BibEmph{Euler H., Kockel B.} {The scattering of light by light in
  Dirac\textquoteright{}s theory}~//
  \href{http://dx.doi.org/10.1007/BF01493898}{Naturwiss.} \BibDash
\newblock 1935. \BibDash
\newblock V.~23, no.~15. \BibDash
\newblock P.~246--247.

\bibitem{Heisenberg:1935qt}
\selectlanguageifdefined{english}
\BibEmph{Heisenberg W., Euler H.} {Consequences of Dirac's theory of
  positrons}~// \href{http://dx.doi.org/10.1007/BF01343663}{Z. Phys.} \BibDash
\newblock 1936. \BibDash
\newblock V.~98, no. 11-12. \BibDash
\newblock P.~714--732. \BibDash
\newblock arXiv:physics/0605038.

\bibitem{Brodin:2001zz}
\selectlanguageifdefined{english}
\BibEmph{Brodin G., Marklund M., Stenflo L.} {Proposal for Detection of QED
  Vacuum Nonlinearities in Maxwell's Equations by the Use of Waveguides}~//
  \href{http://dx.doi.org/10.1103/PhysRevLett.87.171801}{Phys. Rev. Lett.}
  \BibDash
\newblock 2001. \BibDash
\newblock V.~87. \BibDash
\newblock P.~171801. \BibDash
\newblock arXiv:physics/0108022.

\bibitem{Eriksson:2004cz}
\selectlanguageifdefined{english}
\BibEmph{Eriksson D., Brodin G., Marklund M., Stenflo L.} {A Possibility to
  measure elastic photon-photon scattering in vacuum}~//
  \href{http://dx.doi.org/10.1103/PhysRevA.70.013808}{Phys. Rev. A}. \BibDash
\newblock 2004. \BibDash
\newblock V.~70. \BibDash
\newblock P.~013808. \BibDash
\newblock arXiv:physics/0411054.

\bibitem{Bogorad:2019pbu}
\selectlanguageifdefined{english}
\BibEmph{Bogorad Z., Hook A., Kahn Y., Soreq Y.} {Probing Axionlike Particles
  and the Axiverse with Superconducting Radio-Frequency Cavities}~//
  \href{http://dx.doi.org/10.1103/PhysRevLett.123.021801}{Phys. Rev. Lett.}
  \BibDash
\newblock 2019. \BibDash
\newblock V. 123, no.~2. \BibDash
\newblock P.~021801. \BibDash
\newblock arXiv:1902.01418.

\bibitem{Shibata:2020don}
\selectlanguageifdefined{english}
\BibEmph{Shibata K.} {Intrinsic resonant enhancement of light by nonlinear
  vacuum}~// \href{http://dx.doi.org/10.1140/epjd/e2020-10420-1}{Eur. Phys. J.
  D}. \BibDash
\newblock 2020. \BibDash
\newblock V.~74, no.~10. \BibDash
\newblock P.~215.

\bibitem{OurArticle}
\selectlanguageifdefined{english}
\BibEmph{Kopchinskii I., Satunin P.} Resonant generation of electromagnetic
  modes in nonlinear electrodynamics: Classical approach~//
  \href{http://dx.doi.org/10.1103/PhysRevA.105.013508}{Phys. Rev. A}. \BibDash
\newblock 2022. \BibDash 01. \BibDash
\newblock V. 105. \BibDash
\newblock P.~013508. \BibDash
\newblock URL: \BibUrl{https://link.aps.org/doi/10.1103/PhysRevA.105.013508}.

\bibitem{Maxima}
\selectlanguageifdefined{english}
{\url{https://github.com/Ilia-Ko/Supplemental-Materials/tree/main/Nonlinear-ED/Part-I}}.

\bibitem{Hill:2014}
\selectlanguageifdefined{english}
\BibEmph{Hill D.} Electromagnetic Fields in Cavities: Deterministic and
  Statistical Theories~//
  \href{http://dx.doi.org/10.1109/MAP.2014.6821806}{Antennas and Propagation
  Magazine, IEEE}. \BibDash
\newblock 2014. \BibDash 02. \BibDash
\newblock V.~56. \BibDash
\newblock P.~306--306.

\end{thebibliography}

\end{document}